\documentclass[pdflatex,sn-mathphys-num]{sn-jnl}

\usepackage{graphicx}%
\usepackage{multirow}%
\usepackage{amsmath,amssymb,amsfonts}%
\usepackage{amsthm}%
\usepackage{mathrsfs}%
\usepackage[title]{appendix}%
\usepackage{xcolor}%
\usepackage{textcomp}%
\usepackage{manyfoot}%
\usepackage{booktabs}%
\usepackage{algorithm}%
\usepackage{algorithmicx}%
\usepackage{algpseudocode}%
\usepackage{listings}%

\theoremstyle{thmstyleone}%
\theoremstyle{thmstyletwo}%

\theoremstyle{thmstylethree}%
\begin{document}
\title[Article Title]{A Broadband Nanowire Quantum Dot Cavity Design for the Efficient Extraction of Entangled Photons }


\author*[1]{\fnm{Sayan} \sur{Gangopadhyay}}\email{sgangopa@uwaterloo.ca}

\author[2,3]{\fnm{Sasan} \sur{V. Grayli}}

\author[2]{\fnm{Sathursan} \sur{Kokilathasan}}

\author[1,2]{\fnm{Michael} \sur{E. Reimer}}

\affil*[1]{\orgdiv{Institute for Quantum Computing and Department of Physics and Astronomy}, \orgname{University of Waterloo}, \orgaddress{\street{200 University Ave. W}, \city{Waterloo}, \postcode{N2L3G1}, \state{ON}, \country{Canada}}}

\affil*[2]{\orgdiv{Institute for Quantum Computing and Department of Electrical and Computer Engineering}, \orgname{University of Waterloo}, \orgaddress{\street{200 University Ave. W}, \city{Waterloo}, \postcode{N2L3G1}, \state{ON}, \country{Canada}}}

\affil*[3]{\orgdiv{Department of Physics}, \orgname{Simon Fraser University}, \orgaddress{\street{Shrum Science Centre, University Dr W}, \city{Burnaby}, \postcode{V5A 1S6}, \state{BC}, \country{Canada}}}

\abstract{
A bright source of on-demand entangled photons is needed for quantum networks. A single quantum dot in a site-selected nanowire waveguide is a promising candidate for realizing such sources. However, such sources are associated with poor single-photon indistinguishability, limiting their applicability in quantum networks. A common approach for enhancing the single-photon indistinguishability in quantum dot-based entangled photon sources is to implement a broadband optical cavity. Achieving a high-Purcell cavity while retaining the advantages of the nanowire, such as directional emission, a broad operational bandwidth, and high light extraction efficiency, has been a significant challenge. Here, we propose a nanowire cavity based on quasi-bound states in the continuum formed by the strong coupling of two resonant optical modes. We numerically predict this design to support a cavity mode with 4 nm bandwidth and a Purcell enhancement of $\sim$17. This cavity mode enables a directional far-field emission profile (88\% overlap with a Gaussian) with a light extraction efficiency of $\sim$74\%. Our solution opens up a route for generating entangled photon pairs with enhanced extraction efficiency and single-photon indistinguishability for the practical realization of quantum networks.}


\maketitle

\section*{Introduction}\label{sec1}
Bright, indistinguishable, on-demand sources of maximally entangled photons are desired for applications in quantum sensing \cite{Walther}, photonic quantum computing \cite{OBrien2009-bm}, and quantum communication \cite{OBrien2009-bm}. III-V semiconductor quantum dots (QD) in photonic nanostructures have been shown to deterministically emit pairs of photons with near-unity entanglement fidelity via the biexciton (XX) - exciton (X) cascade \cite{pennacchietti2024oscillating, Huber2018-dw}. However, without charge control and Purcell enhancement, these sources can suffer from poor single-photon indistinguishability due to dephasing caused by fluctuating charges near the dot \cite{ yeung2023chip, Reimer2016-ok, Turschmann2019-jb}. One well-established solution to mitigate the detrimental effects of a noisy solid-state environment in these entangled photon sources is to reduce the spontaneous emission lifetime of both the X and XX  by implementing a broadband optical cavity \cite{Liu2019-dz, Bulls, Liu2018-zr}.  

A common approach to integrating QDs in a photonic cavity structure involves top-down fabrication techniques. In this approach, randomly distributed QDs are formed from stress-induced nucleation during the epitaxial growth of semiconductor layers. Then a structure such as a distributed Bragg reflector (DBR) \cite{Somaschi2016-cl} or a Bull's eye \cite{Bulls,Liu2019-dz} is fabricated around the dot.
The high Purcell enhancement in top-etched cavities is dependent on the precise alignment of the QD with the cavity structure \cite{emitter_location}. Uncertainties associated with the optical positioning of QDs can lead to an off-axis placement in the cavity and a reduction in Purcell enhancement, resulting in low device yield \cite{Bulls}. In contrast, the bottom-up approach involving selective-area vapor-liquid-solid (SA-VLS) epitaxy to directly incorporate a single QD in a nanowire (NW) ensures near-perfect on-axis dot alignment \cite{dan}. This enables the potential to grow a nanowire cavity with the dot at its centre. \par

Recent efforts in designing broad bandwidth NW cavities have relied on Fabry-P\'erot resonances that operate in the single-mode regime \cite{kotal, nanopost} to obtain a measured Purcell enhancement of 5 (theoretical Purcell enhancement of 7.9). Metal-cladded nanowire cavities have also been used to combine both a broad bandwidth with high Purcell enhancement ($>$ 60), but come at the expense of a limited extraction efficiency (34\%) and a birefringent cavity with strong polarization splitting \cite{Chellu2024-ja}. In InAsP/InP NW quantum dot sources with a spontaneous emission lifetime of typically $\sim$ 1 ns, a Purcell enhancement in the range of 10 to 30 is preferred. On the other hand, enhancement by 30 times or higher, such as that in the plasmonic NW cavities results in lifetimes lower than 40 ps. Such short lifetimes increase the likelihood of re-excitation from pulsed resonant pumping \cite{Chellu2024-ja}, which would degrade the entanglement fidelity. \par
    Another approach to realizing a broadband cavity in a nanowire with high extraction efficiency that has not been explored is based on bound states in the continuum (BICs). BICs are localized waves that co-exist with the frequency continuum of propagating waves. In photonic structures, such states are formed due to symmetry restrictions or the interference of multiple resonances, forbidding some modes from coupling to propagating fields. A BIC can in principle have an infinitely high Q-factor as a result of perfect localization. A similar phenomenon termed quasi-BIC can be found in isolated high-index dielectric nanoresonators, where the coupling of two resonant modes results in an enhanced Q-factor of one of them. In this work, we introduce a broadband quasi-BIC cavity design for bottom-up NWQDs \cite{Melik-Gaykazyan2021-ww, Hsu2016-fh,friedrich1985interfering} based on a single dielectric NW placed on a metallic mirror. We numerically demonstrate a dielectric NW cavity with a large Purcell enhancement of $\sim$17 and a high extraction efficiency of $\sim$74\% with a directional far-field emission profile. We note that in our cavity design, we are considering a QD source that emits polarization entangled photons in the near-infrared (NIR) via the biexciton-exciton cascade. Typically, in such III-V sources, the X-photon and the XX-photon are separated by $\sim$ 1-2 nm in wavelength. Therefore, we design the spectral range of Purcell enhancement of the cavity to be broader ($\sim$4 nm), facilitating the overlap of the entangled photons with the cavity mode. In addition, we design the cavity to be birefringence free to ensure both polarizations are equally enhanced to maintain the maximally entangled state.
\par In previous work it has been shown that single dielectric resonators can support quasi-BIC modes leading to the formation of high-Q cavities \cite{PRRBIC, MQ, BICPRL}. Despite the high Q-factor, the radiation pattern of these quasi-BIC cavities was shown to be diverging, thereby limiting the collection efficiency using conventional optics.  In our work, we show that the presence of a bottom mirror can transform the far-field radiation pattern from a diverging multi-polar distribution to highly directional at the quasi-BIC resonance. This directional emission profile is a key distinction from a previous study \cite{PRRBIC}, which showed that directional emission from a quasi-BIC cavity embedding a dipole emitter could only be achieved by deliberately moving away from the optimal quasi-BIC condition, i.e., off-resonance where the Purcell enhancement drops drastically.

\section*{Results}\label{sec2}
\subsection*{The optimized quasi-BIC cavity design}

\begin{figure}[H]
\centering
\includegraphics[width=1\textwidth]{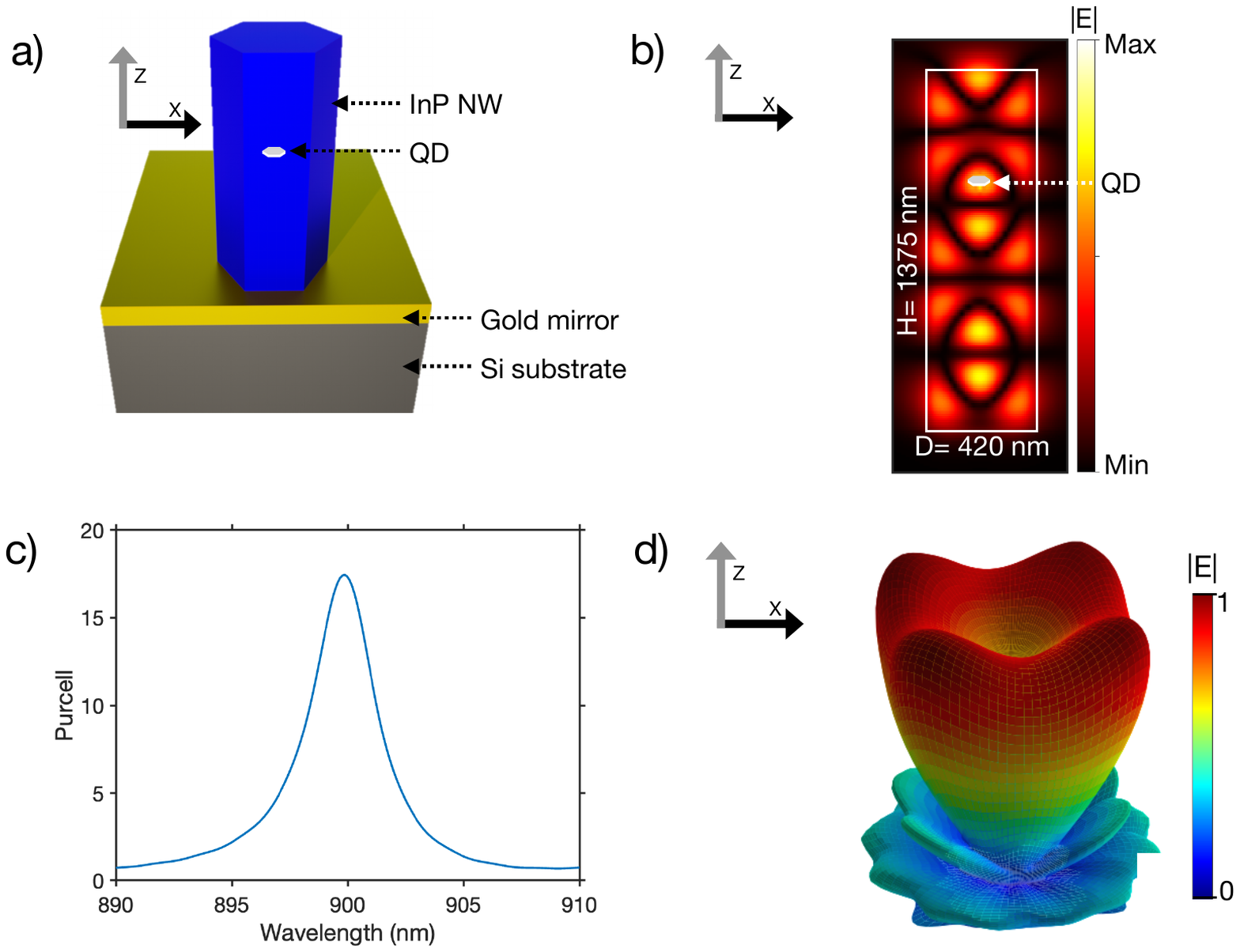}
\caption{(a) Schematic of the quasi-BIC cavity, consisting of a QD in a NW on a gold mirror. (b) E-field profile of the vertical NW cross-section ($\hat{x}-\hat{z}$ plane) with height (H) of 1375 nm and diameter (D) of 420 nm. (c) Simulated Purcell enhancement as a function of wavelength.  (d) Far-field radiation pattern of the NWQD quasi-BIC cavity. }
\label{fig1}
\end{figure}

 In our model of the quasi-BIC cavity, we consider a wurtzite InP NW with a hexagonal cross-section, similar to experimentally investigated NWQD sources of entangled photons \cite{pennacchietti2024oscillating}. A schematic view of the quasi-BIC cavity design is presented in Figure \ref{fig1}a, in which a QD is placed on the NW axis, on top of a gold mirror. To determine the Purcell enhancement of the QD in this structure, we perform Finite-Difference Time-Domain (FDTD) simulations (see Supplementary Information (S1) for details). We modeled the QD by placing a dipole source on the NW axis with an in-plane electric field ($\hat{x}$). In Figure \ref{fig1}b, we show the electric field magnitude ($|E|$) along the vertical cross-section of the NW ($\hat{x}-\hat{z}$ plane). To achieve maximum Purcell enhancement, we position the QD at one of the antinodes of $|E|$. To ensure compatibility with the NWQD growth process and to keep substrate defects distant from the emitter \cite{Reimer2016-ok}, we placed the QD at the second antinode from the top (Figure \ref{fig1}b) in our optimal design. Figure \ref{fig1}c illustrates the Purcell enhancement achieved with the optimal cavity dimensions (height of 1375 nm and width of 420 nm) that we found. For this cavity, a peak Purcell factor of $\sim$17 is obtained for the target wavelength of 900 nm with a full-width at half-maximum (FWHM) of 4 nm. This FWHM of the cavity accommodates the typical wavelength separation of XX and X photons, needed for bright entangled photon pairs.

 Figure \ref{fig1}d shows the far-field emission profile of the QD in the optimized quasi-BIC cavity. This cavity design results in an extraction efficiency of $74\%$, defined as the fraction of total electric field intensity emitted by the QD into the central directional lobe within an NA of 0.8. The far-field emission profile has an overlap of 88\% with a Gaussian, good for fiber coupling.\par
A thin layer of $\mathrm{SiO_2}$ buffer between the NW and the gold has been shown to improve the modal reflectivity of the $HE_{11}$ mode \cite{Claudon2013-wn}. We found that a $12$ nm thick oxide layer optimizes the Purcell enhancement and extraction efficiency when the dipole is placed at the second antinode from the top.  \par

\subsection*{Tunability}
\begin{figure}[H]
\centering
\includegraphics[width=1\textwidth]{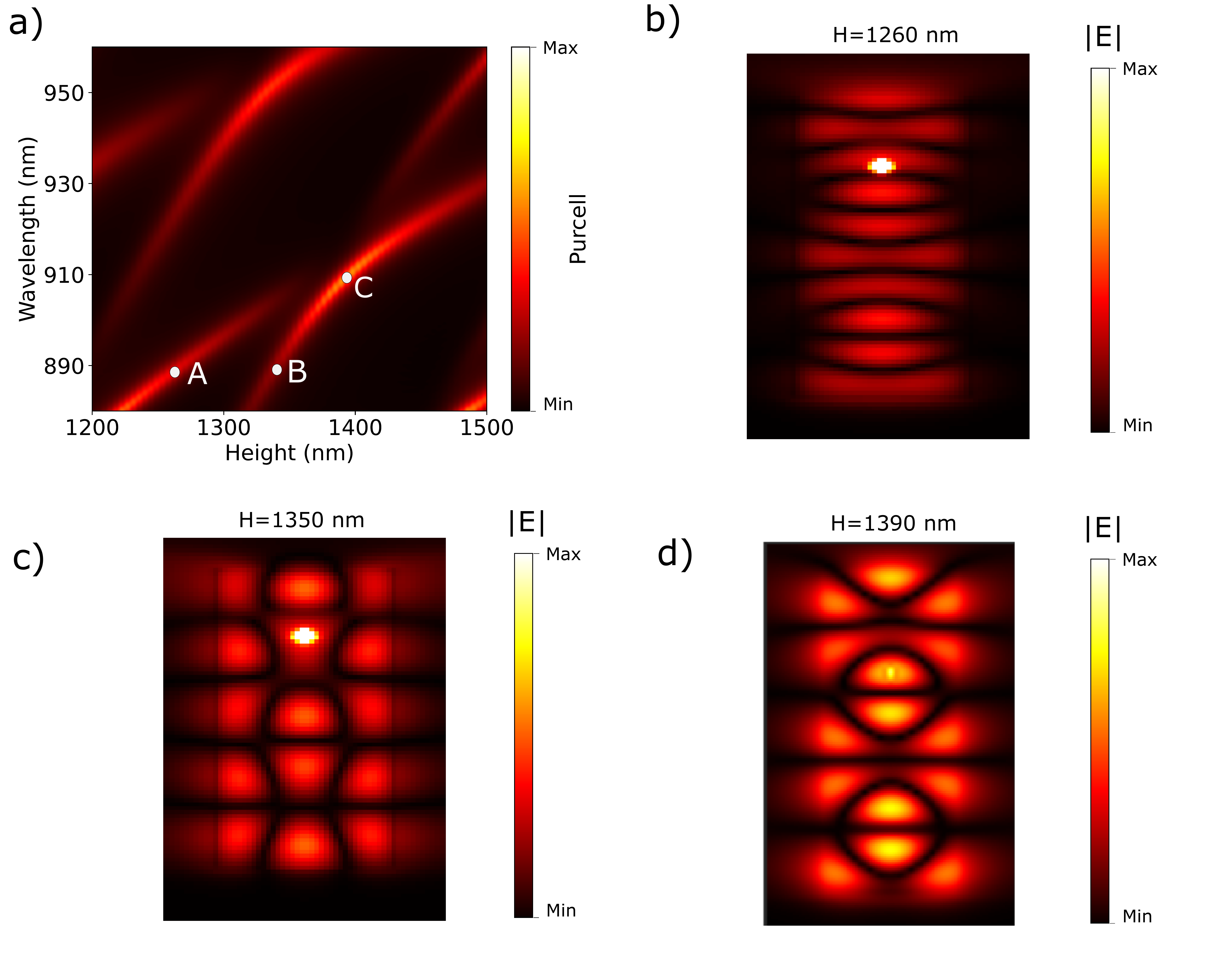}
\caption{(a) Purcell enhancement as a function of wavelength for a nanowire with diameter 420 nm, equipped with a bottom gold mirror and a QD emitter placed on-axis at 30 nm below the top. As the height of the nanowire is increased from 1200 nm to 1500 nm, two resonances labeled A and B undergo strong coupling leading to the formation of an avoided crossing at point C. The maximum Purcell enhancement is obtained at this point C due to the formation of a quasi-BIC. (b), (c) and (d) show the electric field intensity of the resonances labeled A, B and C, respectively at the vertical cross-section of the nanowire.  }\label{fig2}
\end{figure}

To identify the quasi-BIC cavity resonance, we study the Purcell enhancement as a function of wavelength and height. Figure \ref{fig2}a shows two varying resonant modes (labeled A and B) that form an avoided crossing (labeled C) at a height of 1390 nm and wavelength of 910 nm. This phenomenon is a result of the strong coupling of the two modes, A and B \cite{MQ}. To identify these two modes, we perform a modal analysis of an infinitely long InP waveguide with a hexagonal cross-section and a chosen diameter, D = 425 nm. We find that the only two guided modes that are excited by an on-axis quantum dot with an in-plane dipole ($\hat{x}$-direction) are the $HE_{11}$ and the $EH_{11}$ modes (Supplementary Fig. S1).\par
 We attribute the rapidly varying resonant mode (labeled B in Figure \ref{fig2}a) with Fabry-P\'erot-like characteristics (see Figure \ref{fig2}b) to the $HE_{11}$ mode. In contrast, the slowly varying resonance in comparison to B (labeled A in Figure \ref{fig2}a) is ascribed to Mie-like characteristics (see Figure \ref{fig2}c) originates from the $EH_{11}$ mode. At the point of their avoided crossing, the $EH_{11}$ mode vanishes, while the $HE_{11}$ mode is enhanced, as evident in the increase of the Purcell factor in Figure \ref{fig2}a. The electric field of the quasi-BIC cavity mode at the avoided crossing is shown in Figure \ref{fig2}d, highlighting the strong coupling of the two modes A and B.   \par

\begin{figure}[H]
    \centering
    \includegraphics[width=1.0\linewidth]{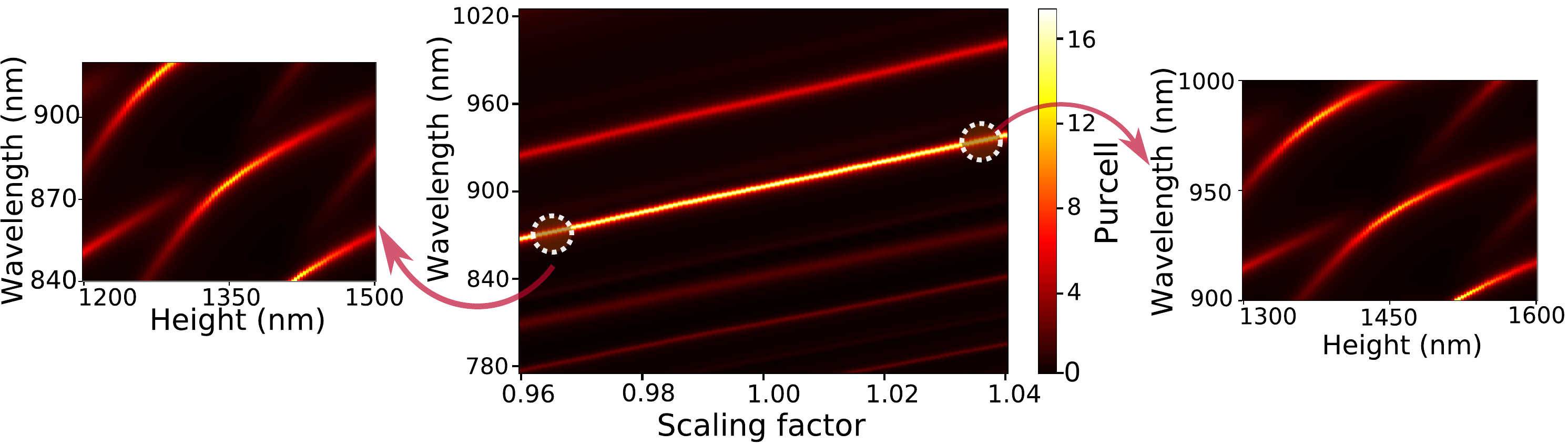}
    \caption{Purcell factor as a function of wavelength and scaling factor ($\mathrm{s}$). The quasi-BIC resonance occurs at the wavelength with the highest Purcell factor. To determine the cavity dimensions for a target wavelength, a corresponding scaling factor (s) can be found from this plot. The cavity dimensions can be calculated as $\mathrm{H'}=\mathrm{s.H}$ and $\mathrm{D'}=\mathrm{s.D}$ where D and H correspond to a design wavelength of 900 nm. The avoided crossings formed at 940 nm and 870 nm  for scaling factors 1.035 and 0.965, respectively (indicated by dashed circles) are shown in the side panels. }
    \label{fig:scaling}
\end{figure}

In Figure \ref{fig:scaling}, we show that the quasi-BIC cavity resonance  ($\lambda$) can be tuned by up to 70 nm by multiplying D and H by a common scaling factor (s) while preserving the aspect ratio (D/H). From the observed trend in Figure \ref{fig:scaling} we calculate $\mathrm{s}=\frac{\lambda'-24.97}{873.66}$. Here, we assume a constant refractive index of the material over the entire wavelength range. In practice, this assumption only holds for a small range of wavelengths, depending on the dispersion relation of the material. For InP the tuning range is $\sim$ 10 - 20 nm. This scaling method provides some wavelength tunability to the cavity design without the need for brute-force parameter sweeps.

\subsection*{Directional emission: role of the mirror}
Without the mirror, the dipole emission pattern diverges from the quasi-BIC NW cavity mode \cite{MQ,PRRBIC}. The emission pattern was made more directional in previous work by adjusting the aspect ratio (D/H) to favor the Fabry-Perot cavity mode; however, the improved directionality came at the cost of reduced Purcell enhancement \cite{PRRBIC, nanopost, Jacobsen2023-el}. To overcome this issue, we introduce a gold mirror at the bottom of the NW base and numerically show that it is possible to achieve both directional emission and high Purcell enhancement from a quasi-BIC cavity. 

\begin{figure}[h]
\centering
\includegraphics[width=1\textwidth]{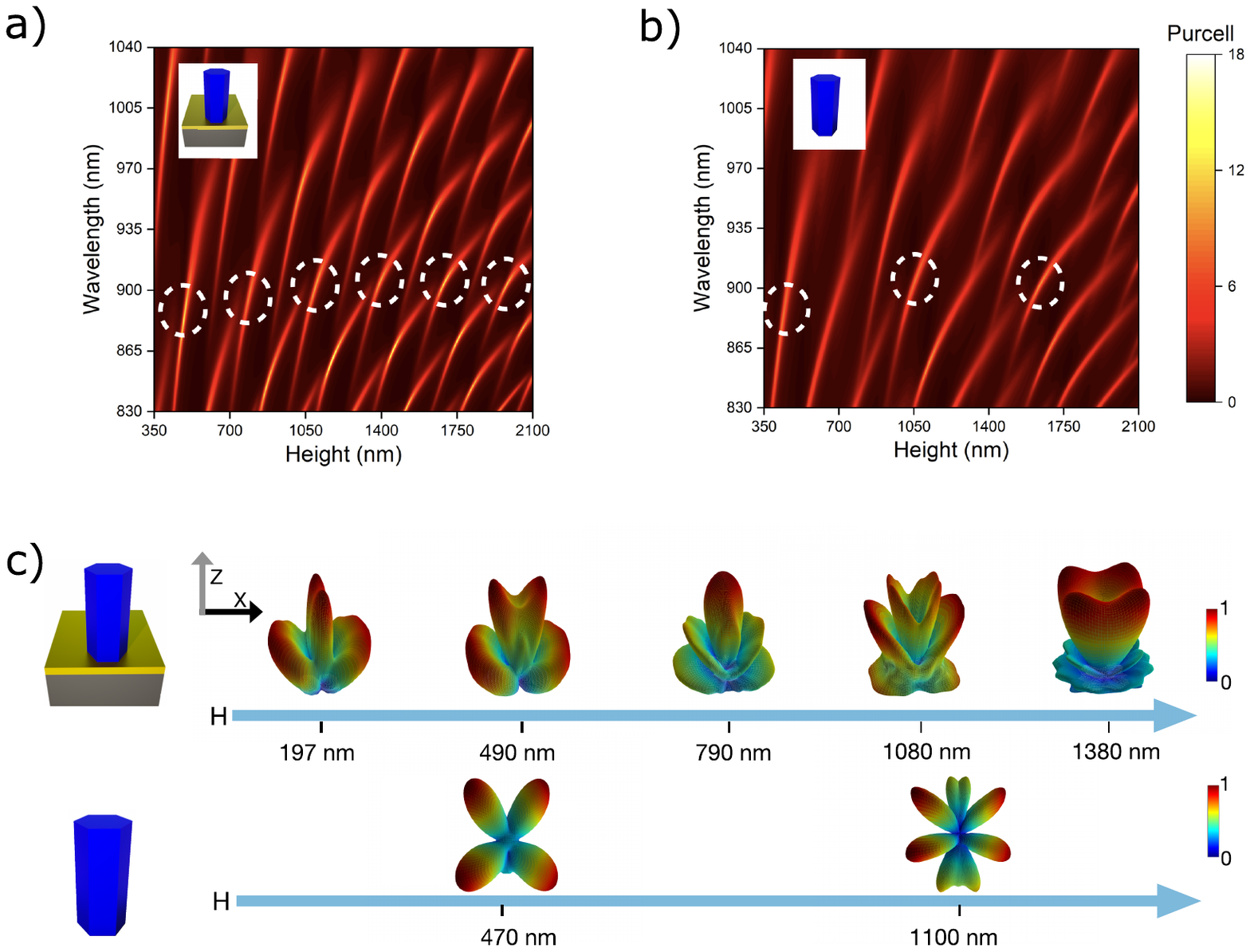}
\caption{ (a) Purcell enhancement as a function of wavelength for a nanowire with D = 420 nm placed on a gold mirror with the QD on the NW axis located 30 nm from the top. As the height of the nanowire is varied from 350 nm to 2.1 $\mu m$, 6 occurrences of the quasi-BIC condition can be observed (dashed white circles). (b) With the same nanowire diameter and QD location, in the absence of the bottom mirror, the number of quasi-BIC occurrences is reduced to 3. (c) Far field radiation profiles of all the quasi-BIC modes as a function of height with and without a bottom gold mirror. The radiation pattern without a bottom mirror corresponds to multipoles of increasing order, starting with a quadrupolar pattern at H=470 nm. However, in the presence of a bottom mirror, the radiation pattern was found to become more convergent with increasing nanowire height. Among all the radiation patterns, the one corresponding to H=1380 nm, is the most directional.}\label{fig3}
\end{figure}

We begin by identifying the quasi-BIC modes for varying NW heights below $2\ \mathrm{\mu m} $ (white circles in Figure \ref{fig3}a,b). We investigate two cases. One case where the NW is placed on a mirror and a second hypothetical case for a NW suspended in air. In the frist case, the bottom mirror enables both even and odd standing waves to be supported by the resonator. As a result, there are twice as many quasi-BIC occurrences close to the design wavelength for a given range of heights for the NW on the mirror (Figure \ref{fig3}a) in comparison to the suspended NW in air (Figure \ref{fig3}b). In Figure \ref{fig3}c, we plot the radiation patterns of quasi-BIC mode occurrences near $\lambda= 900 \ \mathrm{nm}$ in both cases. For the NW in air (bottom row), the radiation pattern starts as a quadrupolar distribution and increases in polarity for increasing height. In this case, the far-field emission lacks directionality with the radiation profile splitting into multiple lobes. In practice, to efficiently collect such  diverging emission, additional optical modules, e.g., adaptive optics, are required. When the NW is placed on a gold mirror (top row), the radiation pattern is transformed into being more directional for some heights (such as H=790 nm and H=1380 nm) as the intensity at their centres is more dominant. We note that the most directional emission, obtained for  H=1380 nm, arises from an odd resonance. This mode has no counterpart in the case without a mirror. Therefore, the presence of a bottom mirror shapes the emission's far-field profile, making it more suitable for free-space collection and subsequent fiber coupling. 

\begin{figure}[h]
\centering
\includegraphics[width=1\textwidth]{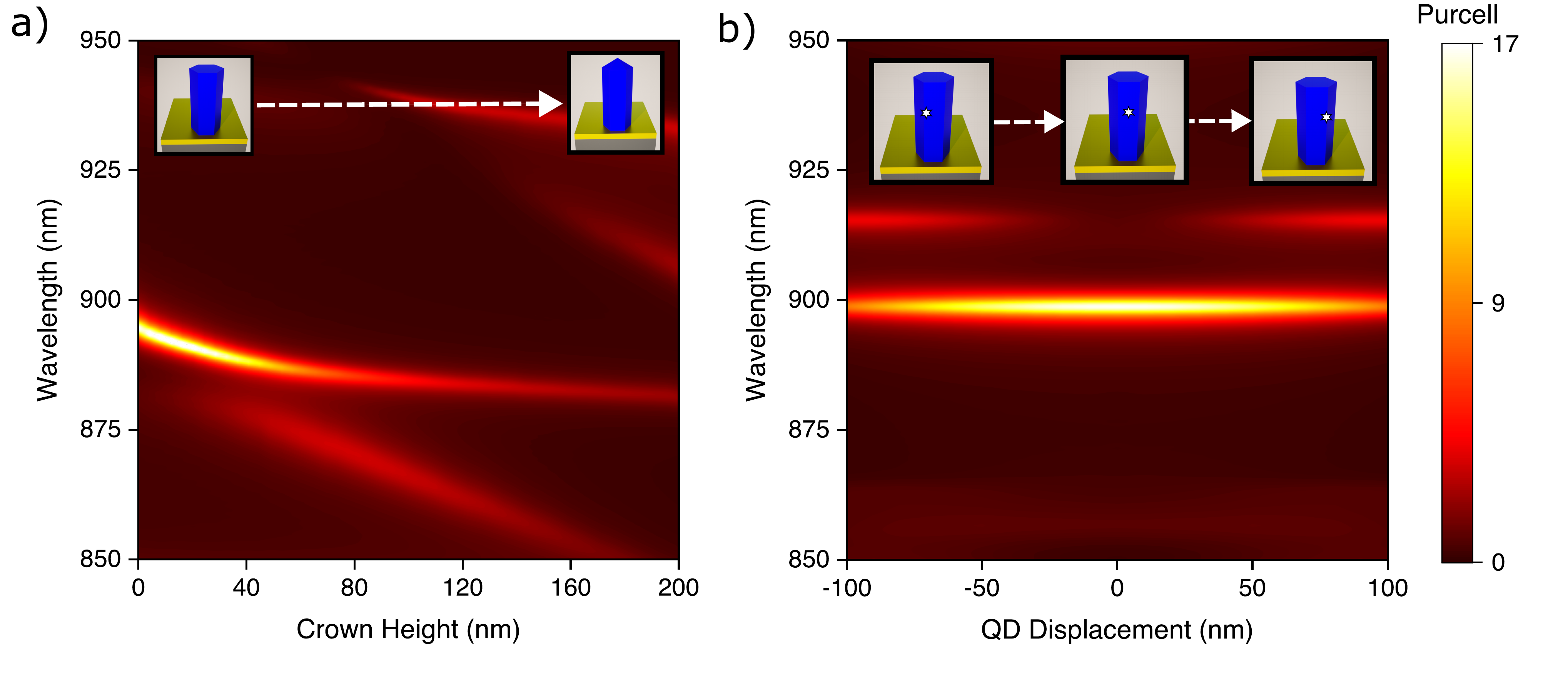}
\caption{(a) Bottom-up nanowire growth often results in a top crown (shown in inset) instead of a flat top. For the optimal dimensions (D=1375 nm and  H= 420 nm), the crown height is varied and the Purcell enhancement is plotted as a function of wavelength and crown height. (b) In top-etched nanowires, there is a significant chance of off-axis QD placement (shown in inset). Purcell enhancement has been plotted against wavelength for QD displacements of up to 100 nm.}\label{fig5}
\end{figure}

\subsection*{Tolerance to fabrication imperfections}

It is difficult to achieve a perfectly flat top facet of the NW with the bottom-up growth process \cite{dalacu2012ultraclean}. There ends up being a conical `crown' at the top instead of a flat facet (inset of Figure \ref{fig5}a). In Figure \ref{fig5}a, we plot the Purcell enhancement as a function of wavelength for crown heights up to 200 nm. We find that the cavity provides a Purcell enhancement greater than 10 for crown heights smaller than 90 nm, which have been successfully grown. To demonstrate the suitability of our cavity design for self-assembled quantum dots where precise on-axis positioning may not be possible, we study the effect of off-axis QD placement on the Purcell enhancement. In Figure \ref{fig5}b, we plot the Purcell factor as a function of wavelength for quantum dot displacements of up to 100 nm from the axis. We find that the peak Purcell enhancement is higher than 10 for a displacement of 100 nm from the cavity-axis. 

\subsection*{Summary}
We have designed a quasi-BIC cavity based on a single dielectric nanoresonator to simultaneously offer a high Purcell enhancement of 17, with a relatively broad bandwidth of 4 nm and 74\% extraction efficiency into a directional far-field profile. The broadband nature of our designed cavity is of particular interest for quantum dot entangled photon sources as the 4 nm bandwidth is sufficient to enhance both the X and XX photons. Typically, they are separated by $\Delta\lambda\approx1-2$ nm due to the biexciton binding energy.\par 

The mechanism of quasi-BIC mode formation by the strong coupling of two modes in dielectric NWs has been well studied, but there has been limited focus on efficiently extracting light from them. In this work, we have analyzed the effect of integrating a gold bottom mirror with a quasi-BIC NW cavity and discovered that it is possible to have a high extraction efficiency without sacrificing the Purcell enhancement. Our work also provides a systematic approach for designing such high Purcell cavities with the intended application to quantum light sources. While this work was motivated with site-selected, epitaxial NWQD sources, the results are platform agnostic. Finally, the experimental realization of our design can be used for high rate entangled photon sources for entanglement swapping - a key milestone towards realizing quantum networks.

\noindent

\backmatter

\bmhead{Acknowledgements}
We acknowledge funding from 
National Sciences and Engineering
Research Council of Canada (NSERC), Mitacs, and National Research Council Canada's Internet of Things: Quantum Sensors Challenge program (QSP) to support this research. 

\bmhead{Author contribution}
SG and SVG conceived the idea and performed the numerical FDTD calculations under the supervision of MER. SK performed the modal analysis. SG, SVG and MER wrote the manuscript.   

\bmhead{Disclosures}
The authors declare no conflicts of interest.
\noindent
\bmhead{Supplemental document}
\section*{S1. FDTD calculations}
The numerical calculations were performed using FDTD: 3D Electromagnetic Simulator \cite{Lumerical} - a commercially available Maxwell's equation solver. The nanowire geometry was modeled using an extruded polygon object. A dielectric with an index of refraction n=3.44 was used to simulate the nanowire material (Wurtzite InP). For the Gold mirror, Palik's experimental permittivity data was used. A mesh size of 10 nm in all three co-ordinates was used for the nanowire object. For the rest of the simulation area a mesh order of 2 was used. \par
A dipole source was used to model the QD emitter. The Purcell factor for a given frequency can be directly accessed from the results tab of the dipole after running the simulation. \par
For calculating the radiation patterns, a box monitor was used. The box enclosed the nanowire and was placed just above the substrate. The bottom facet of the box was disabled such that all the 5 remaining facets are in the same medium, i.e., air. This ensures that an accurate far-field projection can be computed.  The power data from the box monitor was used for calculating the collection efficiency. We note that special care needs to be taken when selecting the box monitor size. Near-field effects can result in inaccurate far-field projections if the monitors are too close to the structure. A box monitor with a side length of $4\, \mu m$ was used in our calculations.

\section*{S2. Modal Ananlysis}
Modal analysis was performed on an infinitely long circular waveguide by solving fully vectorial Maxwell’s equations. A MATLAB package called Optical Fibre Toolbox \cite{karapetyan2025optical} was used for the calculations. Each mode supported by the structure is associated with a distribution in space for electric and magnetic fields \cite{snyder1983optical}. Once a mode has been calculated, the effective refractive index ($\mathrm{N_{eff}}$) of each mode for a given diameter can also be calculated. Numerical sweeps are performed over the selected diameter range and plotted against the calculated $\mathrm{N_{eff}}$ values for each guided mode (modal dispersion plots). The refractive index of the fiber core and cladding materials in this analysis were InP and Air, with the InP refractive index model based on the handbook by Prof. Edward D. Palik \cite{palik1985handbook}. A scaling factor of 1.14 was applied to the diameter as the nanowire has a hexagonal cross-section.  
\begin{figure}[H]
\centering
\includegraphics[width=0.8\textwidth]{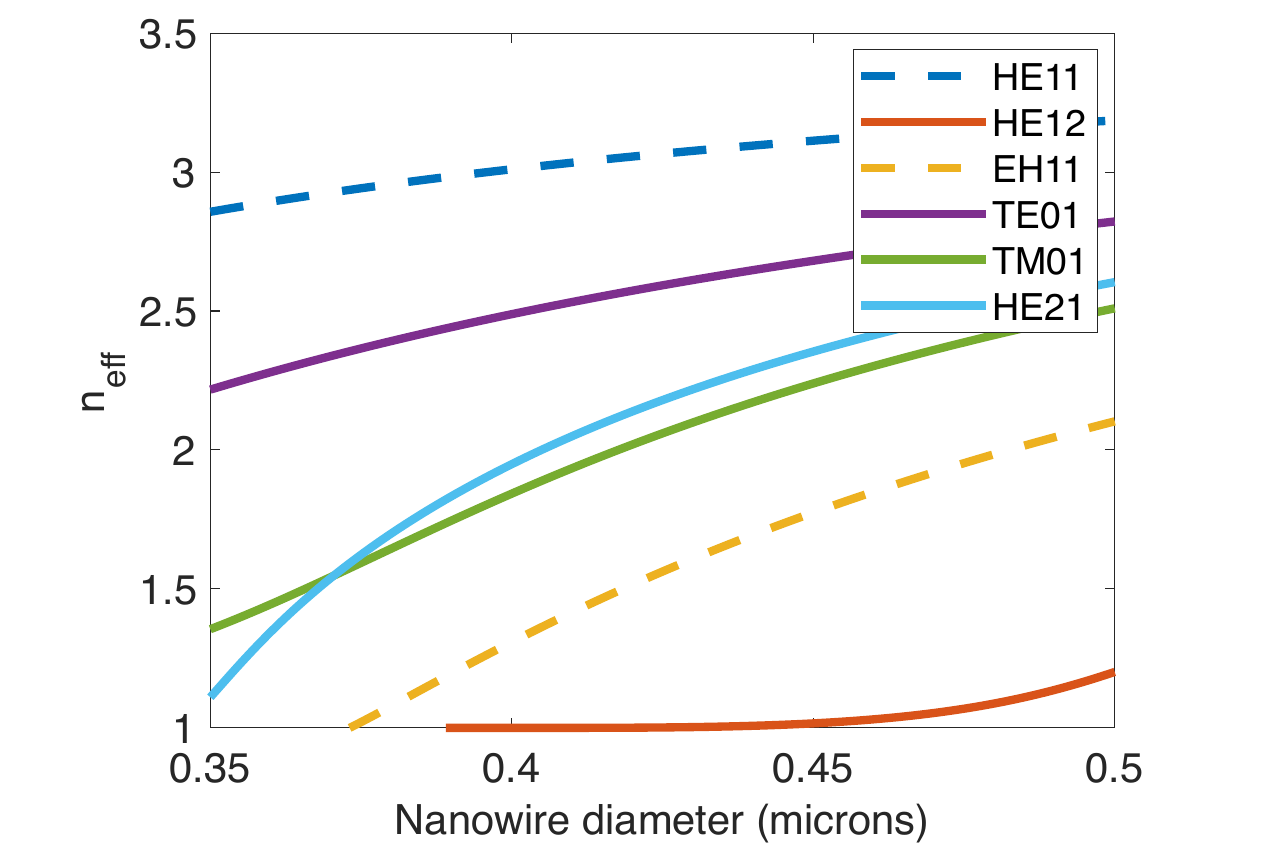}
\caption{Supported modes and their effective refractive indices as a function of the nanowire diameter. The diameter of a hexagonal nanowire is related to a cylindrical nanowire by 2s=1.14D.  }\label{fig_S1}
\end{figure}

\bigskip
\begin{flushleft}%

\bigskip\noindent

\bigskip\noindent

\bigskip\noindent

\bigskip\noindent
\end{flushleft}

\bibliography{sn-bibliography}

\end{document}